\documentclass[fleqn,usenatbib]{mnras}
\pdfoutput=1 
\usepackage[T1]{fontenc}
\DeclareRobustCommand{\VAN}[3]{#2}
\let\VANthebibliography\thebibliography
\def\thebibliography{\DeclareRobustCommand{\VAN}[3]{##3}\VANthebibliography}

\usepackage{graphicx}	
\usepackage{amsmath}	
\usepackage{amssymb}	

\def\NII  {[N{\small{II}}]}
\def\CII  {[C{\small{II}}]}
\def\HII  {H{\small{II}}}
\def\OI  {[O{\small{I}}]}
\def\OIII {[O{\small{III}}]}

\def\micron {$\mu$m}

\title[]{The Case for Thermalization as a Contributor to the \CII\ Deficit}

\author[Sutter et al.]{
Jessica Sutter,$^{1}$\thanks{E-mail: jsutter4@uwyo.edu}
Daniel A. Dale$^{1}$,
Karin Sandstrom$^{2}$,
J.D.T. Smith$^{3}$,
Alberto Bolatto$^{4}$,
Mederic Boquien$^{5}$, 
\newauthor
Daniela Calzetti$^{6}$,
Kevin V. Croxall $^{7}$,
Ilse De Looze$^{8}$,
Maude Galametz $^{9}$,
Brent A. Groves$^{10, 11}$,
George Helou$^{12}$,
\newauthor
Rodrigo Herrera-Camus$^{13}$,
Leslie K. Hunt$^{14}$,
Robert C. Kennicutt$^{15, 16}$,
Eric W. Pelligrini $^{17}$,
Christine Wilson$^{18}$,
\newauthor
Mark G. Wolfire$^{4}$
\\
$^{1}$Department of Physics \& Astronomy, University of Wyoming, Laramie WY \\
$^{2}$Center for Astrophysics and Space Sciences, Department of Physics, University of California, San Diego, CA \\
$^{3}$Department of Physics \& Astronomy, University of Toledo, Toledo, OH \\
$^{4}$Department of Astronomy, University of Maryland, College Park, MD \\
$^{5}$Centro de Astronom\'ia (CITEVA), Universidad de Antofagasta, Avenida Angamos 601, Antofagasta, Chile \\
$^{6}$Department of Astronomy, University of Massachusetts, Amherst MA \\
$^{7}$Expeed Software, Columbus, OH \\
$^{8}$Department of Physics \& Astronomy, University College London, London, UK \\
$^{9}$Laboratoire AIM, CEA, Universit\'e Paris Diderot, IRFU/Service d’Astrophysique, Gif-sur-Yvette, France \\
$^{10}$Research School of Astronomy \& Astrophysics, Australian National University, Canberra, Australia \\
${11}$  University of Western Australia, Perth Australia \\
$^{12}$IPAC, California Institute of Technology, Pasadena, CA \\
$^{13}$Departmento de Astronom\'ia, Facultad de Ciencias F\'isicas y Matem\'aticas, Universidad de Concepci\'on, Concepci\'on, Chile \\
$^{14}$INAF -  Osservatorio Astrofisico di Arcetri, Firenze, Italy \\
$^{15}$Steward Observatory, University of Arizona, Tucson, AZ\\
$^{16}$George P. and Cynthia W. Mitchell Institute for Fundamental Physics and Astronomy, Texas A\&M University, College Station, TX \\
$^{17}$Institute for Theoretical Astrophysics Heidelberg, Germany \\
$^{18}$Department of Physics \& Astronomy, McMaster University, Hamilton, ON Canada
}

\date{Accepted Feb 17th, 2021. Received YYY; in original form ZZZ}

\pubyear{2021}

\begin{document}
\label{firstpage}
\pagerange{\pageref{firstpage}--\pageref{lastpage}}
\maketitle

\begin{abstract}
The \CII\ deficit, which describes the observed decrease in the ratio of \CII~158~\micron\ emission to continuum infrared emission in galaxies with high star formation surface densities, places a significant challenge to the interpretation of \CII\ detections from across the observable universe.  In an attempt to further decode the cause of the \CII\ deficit, the \CII\ and dust continuum emission from 18 Local Volume galaxies has been split based on conditions within the interstellar medium where it originated.  This is completed using the Key Insights in Nearby Galaxies: a Far--Infrared Survey with \textit{Herschel} (KINGFISH) and Beyond the Peak (BtP) surveys and the wide--range of wavelength information, from UV to far--infrared emission lines, available for a selection of star--forming regions within these samples.  By comparing these subdivided \CII\ emissions to isolated infrared emission and other properties, we find that the thermalization (collisional de-excitation) of the \CII\ line in \HII\ regions plays a significant role in the deficit observed in our sample. 
\end{abstract}




\section{Introduction}
The \CII~158~\micron\ line is an increasingly alluring tool for understanding how galaxies evolve over time.  As one of the major cooling channels for the interstellar medium \citep[ISM,][]{Wolfire2003}, \CII~158~\micron\ emission can represent up to a few percent of the total infrared output of a star--forming galaxy \citep{Malhotra2001, Smith2017}.   The far--infrared wavelength of the \CII\ line make it a frequent target of high--redshift surveys, as it falls within the observable bands of millimeter arrays across a wide range of redshifts ($2<z<7$) \citep[e.g.][]{Rybak2019, Ginolfi2020}.  Except in extreme cases, the long wavelength of the \CII\ line also allows it to pass through gas and dust with little to no attenuation, which, along with \CII\ line's role as a major ISM coolant, frequently make it the brightest observed emission line in star--forming galaxies \citep{Luhman2003}.  

In order to interpret the multitude of new high--$z$ \CII\ detections from millimeter observatories like The Atacama Large Millimeter Array (ALMA) \citep[e.g][]{Schaerer2020}, it is important that we have a thorough understanding of the production of the \CII\ line in well--studied, Local Volume galaxies.  Previous works have suggested that \CII\ could trace star formation rate (SFR) \citep{DeLooze2011,DeLooze2014, HerreraCamus2015}, shock--heated gas \citep{Appleton2017}, or diagnose properties of the atomic interstellar medium \citep{Contursi2002, HerreraCamus2017}.  Before these potential uses of the \CII\ line can be applied to high--z \CII\ measurements, the limitations of the \CII\ line must be understood.

A complication discovered in this pursuit is the \CII\ deficit.  The \CII\ deficit describes the decreasing trend in the ratio of \CII~158~\micron\ luminosity to the total infrared luminosity (TIR) from 5--1100 \micron\ with increasing star--formation rates or hotter dust \citep{Malhotra2001, Smith2007, Sargsyan2012, Sutter2019}.  This effect has been measured in a wide variety of galaxies, including Ultra--Luminous Infrared Galaxies (ULIRGs) \citep{DiazSantos2017, HerreraCamus2018a} and high--$z$ galaxies \citep{Capak2015, Decarli2018}.  The \CII\ deficit is especially detrimental when using the \CII\ line as a measure of star formation rate (SFR), since the TIR luminosity, which indicates absorbed UV/optical emission from young stars re-radiated by dust, is expected to be an accurate SFR tracer as well \citep{DeLooze2014, HerreraCamus2015}.   

Multiple processes have been proposed to explain the cause of the \CII\ deficit.  These include: (1) \CII\ self--absorption or \CII\ becoming optically thick in high density environments \citep{Abel2007, Neri2014}, (2) changes in photoelectric heating efficiency caused by the charging of polycyclic aromatic hydrocarbons (PAHs) \citep{Malhotra2001, Croxall2012}, (3) the conversion of singly ionized carbon to doubly--ionized carbon in AGN--host galaxies \citep{Langer2015}, (4) thermalization of the \CII\ line in high--density \HII\ regions \citep{DiazSantos2017}, (5) \OI\ and other far--infrared fine structure lines overcoming \CII\ as the dominant cooling sources \citep{Brauher2008}, (6) an increased ionization parameter in dense galaxies leading to a greater fraction of FUV radiation being absorbed by dust and therefore less availability of carbon--ionizing radiation  \citep{DeLooze2014, HerreraCamus2018a}, (7) increasing G$_0$/n values within a single PDR are expected to decrease \CII/TIR values through grain charging and \OI\ becoming the dominant cooling line \citep{Kaufman1999}, or some combination of these effects.  Determining the precise cause of the \CII\ deficit will help establish the utility of measurements of this line in the high--redshift universe.  

To study the behavior of the \CII\ deficit, it is important to consider the different environments in which C$^+$ can be found.  As carbon has a relatively low ionization potential of 11.3~eV (110~nm), lower than the 13.6~eV necessary to ionize hydrogen, C$^+$ is present in many ISM phases, complicating the determination of the cause of the deficit.  

This paper builds on the results of \citet{Sutter2019}, which measured the [CII] deficit in a sample of nearby, star--forming galaxies and found the deficit has a more significant effect in the ionized phases of the ISM (i.e. phases where hydrogen is ionized) than in neutral phases.  This was found by measuring the clear differences in the trends observed in the [CII]/TIR values from the isolated ionized and neutral phases as a function of dust temperature (see Figure 5, \citet{Sutter2019}).  In order to elucidate the cause of this difference, we present the results from a novel analysis of the \CII\ deficit.  Employing the wide range of available photometric observations and spectral emission line detections for this sample, we are able to separate many of the observed fluxes of these galaxies into the phases from which they arose.  By isolating emission from separated phases of the ISM, we are able to isolate the effects of potential causes of the \CII\ deficit.  Preliminary results from this analysis suggest that the thermalization of the \CII\ line in \HII\ regions may play a significant role in the \CII\ deficit across our sample.  

Thermalization occurs when the critical density of a collisional interaction is surpassed.  Below the critical density, line intensities increase with increasing density, $I($line$) \propto n^2$, as higher densities lead to a higher frequency of collisional excitations.  Above the critical density, or the density at which collisional de-excitations and spontaneous radiative decays are equal, line intensities will linearly increase with increasing density ($I($line$) \propto n$) as local thermodynamic equilibrium conditions are approached. It is important to consider thermalization when studying the \CII\ deficit as when SFR and thus TIR luminosity increase, the thermal pressure increases leading to an increased density \citep{Ostriker2010, HerreraCamus2017},  while the \CII\ line would increase comparatively more slowly.  By comparing the density in isolated ISM phases to the \CII~/~TIR ratio for these phases, this theoretical cause of the \CII\ deficit can be observationally tested.

This paper is organized as follows.  Section~\ref{sec:obs} describes the sample and observations used to complete this analysis.  Section~\ref{sec:sep} explains the methods applied to determine different properties and measurements separated by ISM phase.  Finally, Section~\ref{sec:con} lays out the initial conclusions obtained in this work.

\section{Data and Sample}
\label{sec:obs}
This work uses the subset of the Key Insights in Nearby Galaxies: a Far Infrared Survey with \textit{Herschel}, or KINGFISH, sample with \CII~158~\micron, \NII~122~\micron\ and 205~\micron, and \OI~63~\micron\ line detections. The KINGFISH survey includes 61 Local Volume galaxies with imaging and spectroscopy across the infrared spectrum \citep{Kennicutt2011}.  Of the 61 total galaxies, star--forming regions in 18 local galaxies ($D < 20$~Mpc) have the full suite of line observations required for this work.  All line measurements were performed using the Photoconductor Array Camera and Spectrometer (PACS) on the \textit{Herschel Space Telescope} \citep{Poglitsch2010}.  12 of these galaxies were also included in the Beyond the Peak (BtP) survey.  This survey used the Spectral and Photometric Imaging Receiver \citep[SPIRE][]{Griffin2010} to obtain maps of the \NII~205~\micron\ over a wider field of view than the KINGFISH observations, extending the coverage of these galaxies to the more quiescent areas surrounding the star--forming regions targeted at 205~\micron\ by the PACS instrument. The combination of the KINGFISH and BtP surveys cover a total of 120 individual regions in these 18 galaxies.  All line maps were smoothed to the 14\farcs5 point spread function (PSF) of the \NII~205~\micron\ maps using a Gaussian profile.  For the regions in this sample, the 14\farcs5 radius translates to a physical length of 200--2100~pc.  The regions in these two samples have SFR $\sim$ 0.082--3.92 $M_{\odot}$ yr$^{-1}$, stellar masses of log$_{10}$(M$_{\odot}$) $\sim$ 8.97--10.84 and cover log$_{10}$(O/H)$+12 \sim $ 8.1-8.7, making them all normal, star-forming galaxies \citep{Hunt2019, Kennicutt2011}.  Further information about the data processing and this sample can be found in \cite{Sutter2019}. 

In addition to the far--infrared line maps, the KINGFISH survey imaged each galaxy at 70, 100, and 160~\micron\ using the PACS instrument and at 250~\micron\ using the SPIRE instrument on board \textit{Herschel} (the KINGFISH 350~\micron\ and 500~\micron\ data are not utilized in this work).  All 18 galaxies in this study were also included in the \textit{Spitzer} Infrared Nearby Galaxies Survey, or SINGS \citep{Kennicutt2003}.  As part of this survey, each galaxy was imaged at 3.6, 4.5, 5.6, 8.0, and 24~\micron\ using the Infrared Array Camera (IRAC) and the Multi-band Imagine Photometer (MIPS) on the \textit{Spitzer Space Telescope} \citep{Rieke2004}.  These infrared bands primarily measure the warm dust and light from old stars in these galaxies.  Additionally, $ugriz$ data were obtained from the Sloan Digital Sky Survey (SDSS) for each galaxy in this sample. Foreground stars and background galaxies are removed from the SDSS maps.  Further descriptions of the $ugriz$ data processing can be found in \cite{Cook2014}.  To expand the range of photometric data, \textit{GALEX} FUV (1350--1750~\AA) and NUV (1750--2800~\AA) images were also obtained.  All galaxies were imaged with \textit{GALEX} as part of either the Nearby Galaxy Survey (NGS) or the All-Sky Imaging Survey (AIS) \citep{GildePaz2005}.  In order to perform consistent analysis across this wide--range of wavelengths, all imaging data were smoothed to the 250~\micron\ PSF of 18\arcsec\ using a Gaussian profile.

\section{Separation of Ionized and Neutral ISM}
\label{sec:sep}
In order to perform this analysis, the ISM within each galaxy was split into two phases.  The first phase is the ionized phase, characterized by environments where hydrogen is predominately ionized, and includes \HII\ regions and the diffuse ionized ISM.  The second phase is the neutral phase, where hydrogen is predominately neutral, and includes photodissociation regions (PDRs) as well as molecular clouds.  This work proposes a novel method to examine the \CII\ deficit by separating not only the \CII\ emission but also the TIR luminosity and gas properties by ISM phase in unresolved star--forming regions.  By isolating the two phases, the precise causes of the deficit can be tested with fewer complicating factors.

\subsection{Isolating the Ionized vs. Neutral \CII\ Emission}
\begin{figure}[H]
\includegraphics[width=\linewidth]{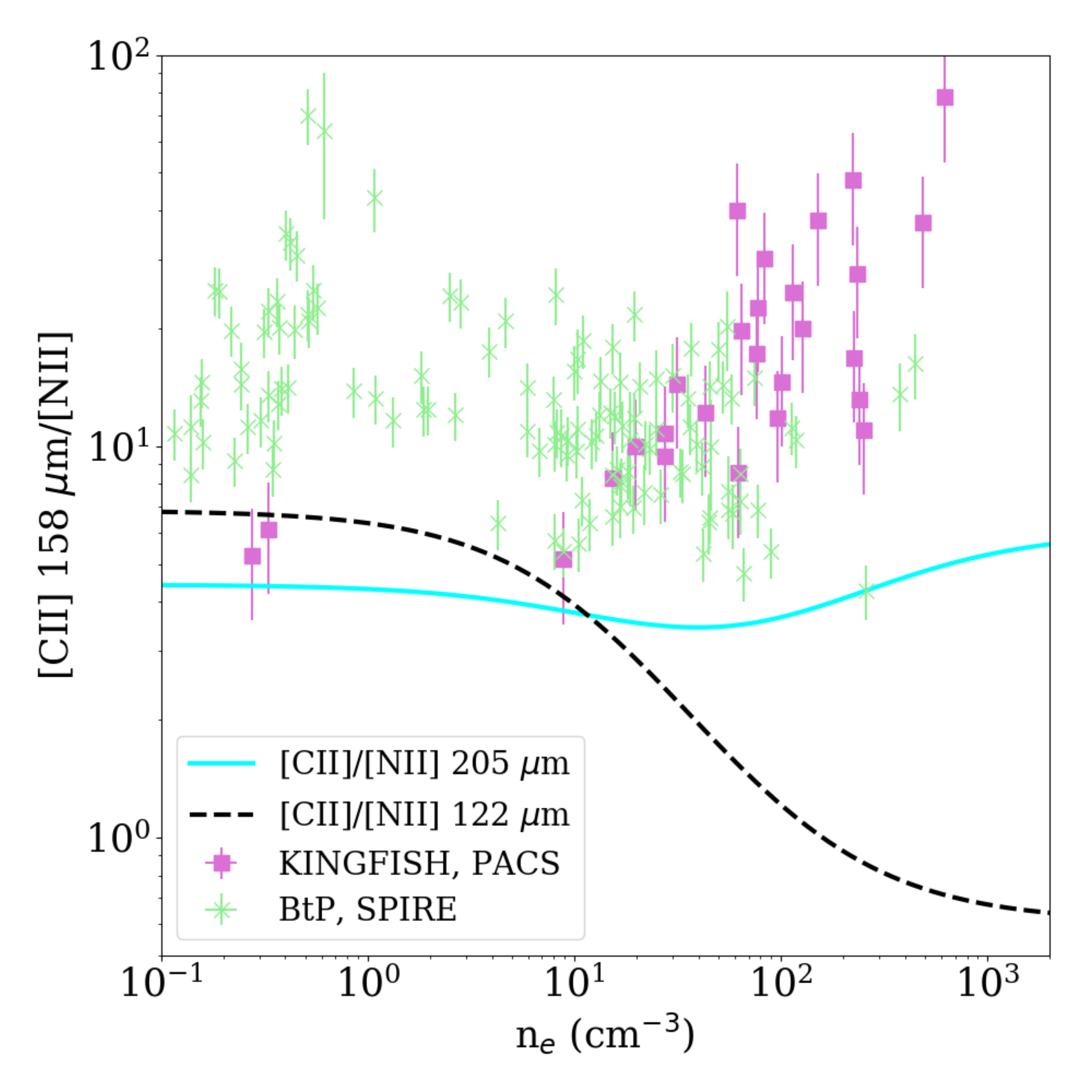}
\caption{The theoretical ratio of the \CII~158~\micron\ line and \NII~122~\micron\ and 205~\micron\ lines for ionized gas are represented by the black dashed line and solid cyan line respectively.  These ratios are used to determine the fraction of the \CII\ emission that arises in the ionized phases of the ISM, where both C$^+$ and N$^+$ are present.  The observed ratio of  \CII~158~\micron\ to the \NII~205~\micron\ emission is shown for each region in our sample.  As all the observations fall above the cyan line, much of the \CII\ emission must originate from the neutral ISM, where no N$^+$ is present.  Theoretical ratios are determined using the method described in \citet{Croxall2012} and are based on Galactic carbon and nitrogen abundances.}
\label{fig:rion}
\end{figure}

The \CII\ emission was divided by ionized and neutral ISM phases using the theoretical relationship between the \CII~158~\micron\ line and the \NII~205~\micron\ line in ionized gas.  Figure~\ref{fig:rion} shows the predicted ratio of the \CII~158~\micron\ line to \NII\ emission from the ionized ISM (\CII/122~\micron\ as the black dashed line, \CII/205~\micron\ as the solid cyan line) as function of electron number density ($n_{\rm{e}}$). While the solid cyan line stays at an approximately constant value of 4, the black dashed line varies by nearly an order of magnitude across typical ISM density conditions.  The measured ratios of \CII~158~\micron\ to the \NII~205~\micron\ emission for the KINGFISH and BtP sample are shown as magenta squares and green crosses, respectively. As nitrogen has an ionization potential of 14.5~eV, well above the 13.6~eV necessary to ionize hydrogen, N$^+$ should primarily exist in the ionized phases of the ISM, so the theoretical ratio of the \CII~/~\NII\ lines can therefore be used to predict the amount of \CII\ emission from ionized phases of the ISM.  The 205~\micron\ line is preferred here over the 122~\micron\ line as the critical densities for collisions with electrons for the 205~\micron\ line and the \CII~158~\micron\ line are fairly similar \citep[$\approx 32$~cm$^{-3}$ for the \NII~205~\micron\ line and $\approx 45$ cm$^{-3}$ for the \CII~158~\micron\ line] []{Oberst2006, Croxall2017}.  This consistency makes the predicted ratio of the 158~\micron\ and 205~\micron\ lines from ionized gas nearly independent of electron number density.  In Figure~\ref{fig:rion}, $n_e$ was determined based on the ratio of the \NII~122~\micron\ and \NII~205~\micron\ lines, as described in Section~\ref{sec:n}.  

As all points lie either above or on the predicted ratio for these two emission lines in ionized gas, we know a large amount of the \CII\ emission for our sample must be from the neutral phases of the ISM, where no ionized nitrogen is present.  We calculate the fraction of the \CII\ emission from the neutral ISM using the equation:
\begin{equation}
f_{\rm{[CII], Neutral}} = \frac{\rm{[CII]}158 - (R_{\rm{Ionized}} \times \rm{[NII]}205)}{\rm{[CII]}158},
\label{eq:fneut}
\end{equation}
where $R_{\rm{Ionized}}$ is the model ratio of the \CII~158~\micron\ to \NII~205~\micron\ emission from the phases of the ISM where both C$^+$ and N$^+$ are present and was determined individually for each of the 120 regions using the measured $n_e$ \citep[cf.][]{Croxall2012}.  For this work, $R_{\rm{Ionized}}$ uses an assumption of a Galactic C/N ratio.  It is possible that changes in the C/N ratio across our sample could have some effect on the measurements of $f_{\rm{[CII], Neutral}}$, but as shown in \citet{Croxall2017}, it is likely that these effects are small compared to variations across the regions in this sample. The method produces measurements of  $f_{\rm{[CII], Neutral}}$ spanning a range from $2\% $ to $94\%$ with  a median value of $66\%$ \citep[see Figure 3,][]{Sutter2019}, similar to what has been found in other works \citep{Croxall2012, Parkin2013, Hughes2015, DiazSantos2017}.

\subsection{Isolating the Ionized vs. Neutral L(TIR)}
The TIR luminosity was determined using the results of \citet{Dale2014}:

\begin{equation}
 L(TIR) = 1.548 \nu L_{\nu}(24 \mu \rm{m} ) + 0.767 \nu L_{\nu}(70 \mu \rm{m}) 
 + 1.285 \nu L_{\nu}(160 \mu \rm{m}) 
\end{equation}

and represents the integrated luminosity from 5--1100~\micron.  We then separate the TIR luminosity arising from the ionized versus neutral ISM phases using the results of the spectral energy distribution (SED) fitting of the 3.6--250~\micron\ thermal dust emission constraints.  The SED fits were completed using the Code Investigating GALaxy Evolution, or CIGALE \citep{Noll2009, Boquien2019} and the full range of photometric observations described in Section~\ref{sec:obs}.  All SED fits were determined using a \citet{BruzualCharlot2003} stellar population, a modified starburst dust attenuation law, and a \citet{Draine2014} dust emission model.  The outputs from the best--fit model along with the \citet{Draine2007} dust models were applied to estimate the fraction of the TIR luminosity from dust heated within the \HII\ regions.  These models include a method for predicting the fraction of dust luminosity heated by an interstellar radiation field (ISRF, modeled as $U$) above a set cutoff value, $U_c$.  By setting $U_c$ to the value of the ISRF expected at the Str\"{o}mgren radius, or the radius of the \HII\ region, we can predict the fraction of dust luminosity coming from the ionized \HII\ region surrounding recent star formation.  This fraction is determined using Equation 18 from \cite{Draine2007}:
\begin{equation}
 f(L_{\rm{dust}}; U > U_c) = \frac{ \gamma \ln(U_{\rm{max}}/U_c) } {(1-\gamma)(1-U_{\rm{min}} / U_{\rm{max}})+\gamma\ln(U_{\rm{max}} / U_{\rm{min}})}
\label{eq:fhii}
\end{equation}

where $\gamma$ is the fraction of the dust luminosity from dust heated by young stars and $U$ is a dimensionless scale factor that when multiplied by the specific energy density of \cite{Mathis1983} quantifies the specific energy density of starlight, i.e. the ISRF.  $U_{\rm{max}}$ is then the maximum value of this scale factor, which is set to $10^6$ for each region \citep[as in][]{Draine2007}, $U_{\rm{min}}$ is the minimum value of this scale factor, which is determined by the CIGALE SED fitting, and $U_c$ is the cutoff value for this scale--factor, which is calculated for each region individually using the definition of the Str\"{o}mgren radius:
\begin{equation}
    \label{eq:stromR}
    R_S = \Big(\frac{4}{3}\pi n_e^2 \alpha N_{\rm Ly} \Big)^{1/3}.
\end{equation}
to determine the expected UV flux within each \HII\ region.  We use an $\alpha = $3.0$ \times 10^{-13}$~cm$^3$~s$^{-1}$ and estimate $N_{\rm Ly}$ by integrating the unattenuated UV flux (i.e. the flux between 91.2--206.6 nm) predicted by the modelled SED.  This allows us to estimate ISRF at the Str\"{o}mgren radius, $U_c$, in Habing units ($1.6\times10^{-3}$~erg~s$^{-1}$~cm$^{-2}$) by dividing the unattenuated UV flux by the surface area of a sphere with the radius determined by Equation~\ref{eq:stromR}.  This gives us a range of $U_c$ from 2.0$\times10^{-1}$--1.7$\times10^4$ with a mean value of 7.93$\times10^2$.  Using this $U_c$ in Equation~\ref{eq:fhii}, we can determine the fraction of the TIR luminosity from within the \HII\ region, or $f_{HII}$.

For the regions included in this sample, $f_{HII}$ ranges from 0.1\%--19\% with a median value of 2.6\%.  The TIR luminosity from the ionized ISM (TIR$_{\rm{Ionized}}$) is $f_{HII} \times L$(TIR) while the neutral TIR luminosity (TIR$_{\rm{Neutral}}$) is the difference between the measured TIR luminosity and TIR$_{\rm{Ionized}}$.

\begin{figure}
\includegraphics[width=\linewidth]{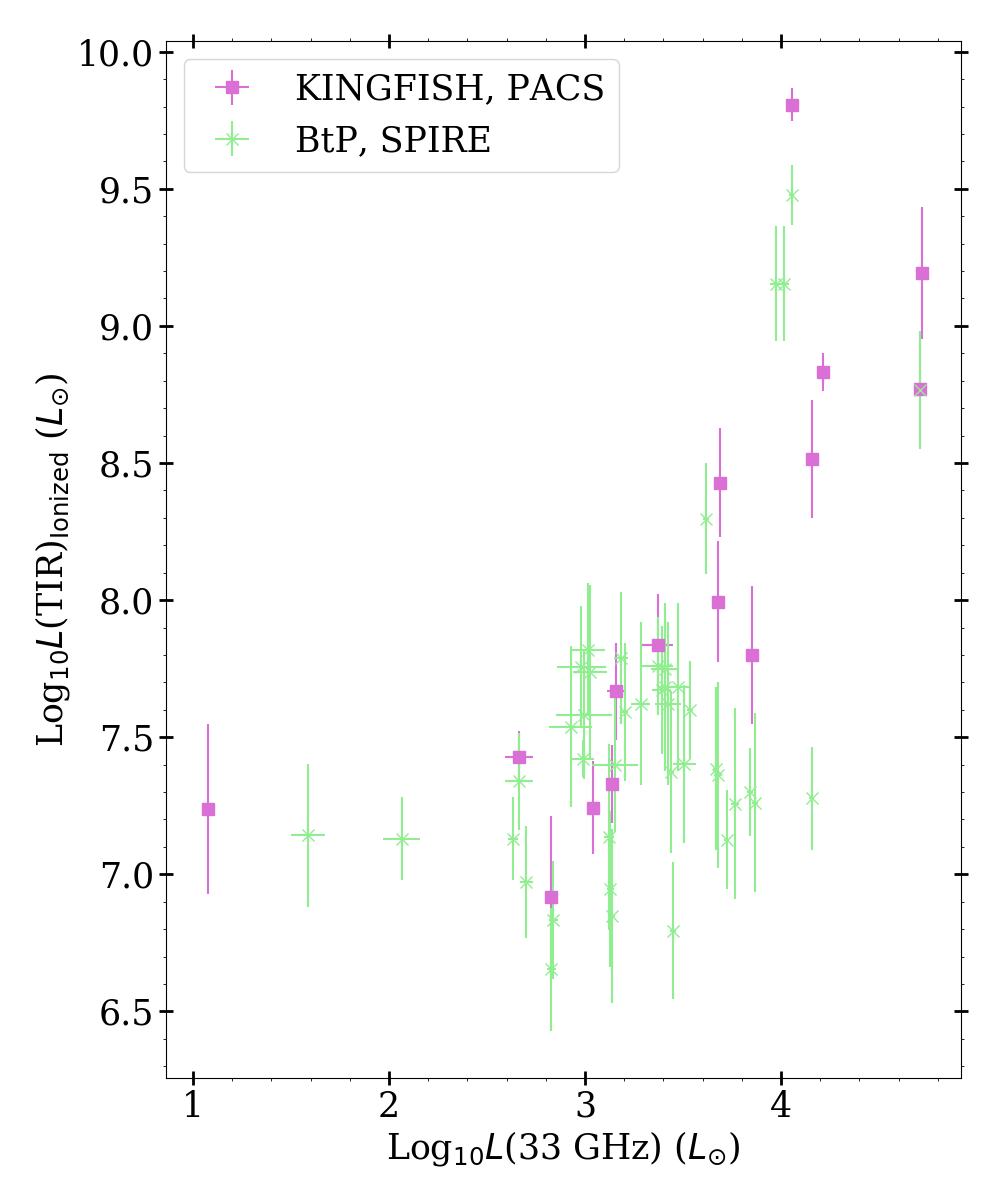}
\caption{Isolated ionized--phase TIR luminosities plotted against the 33~GHz luminosities measured by \citet{Murphy2018} and corrected to include only the thermal fraction of the radio component using the results of \citet{Linden2020} to demonstrate the applicability of using SED models to separate TIR emission.}
\label{fig:tirradio}
\end{figure}

In order to test the viability of this method for isolating the TIR luminosity, measurements of TIR$_{\rm{Ionized}}$ for the subset of regions included in \citet{Murphy2018} are plotted against $L$(33~GHz) in Figure~\ref{fig:tirradio}.  The 33~GHz luminosities were corrected to include only the thermal component using the results of \citet{Linden2020}.  As the thermal component of the 33 GHz luminosity traces free--free emission associated with \HII\ regions \citep{Condon1990, Murphy2011}, it should be well correlated with the TIR emission from \HII\ regions.  As shown in Figure~\ref{fig:tirradio}, there is a clear trend between these two properties, with scatter coming mainly from the BtP regions tracing the more quiescent areas surrounding \HII\ regions.  This suggests that our method for determining TIR$_{\rm{Ionized}}$ works well for star--forming regions, but there may be added uncertainty for the diffuse ISM.

\subsection{Density measurements for the ionized and neutral ISM}
\label{sec:n}
The densities of the neutral and ionized ISM phases were determined separately for each region in this sample.  First, to determine the electron number density within the ionized phases, the ratio of the \NII~122~\micron\ and 205~\micron\ lines was used \citep{HerreraCamus2016}.  As nitrogen has an ionization potential above that of hydrogen, both \NII\ lines should originate in the ionized phases of the ISM.  This ratio has been found to be sensitive to $n_e$ spanning 10--1000 cm$^{-3}$.   

\begin{figure}[H]
\includegraphics[width=\linewidth]{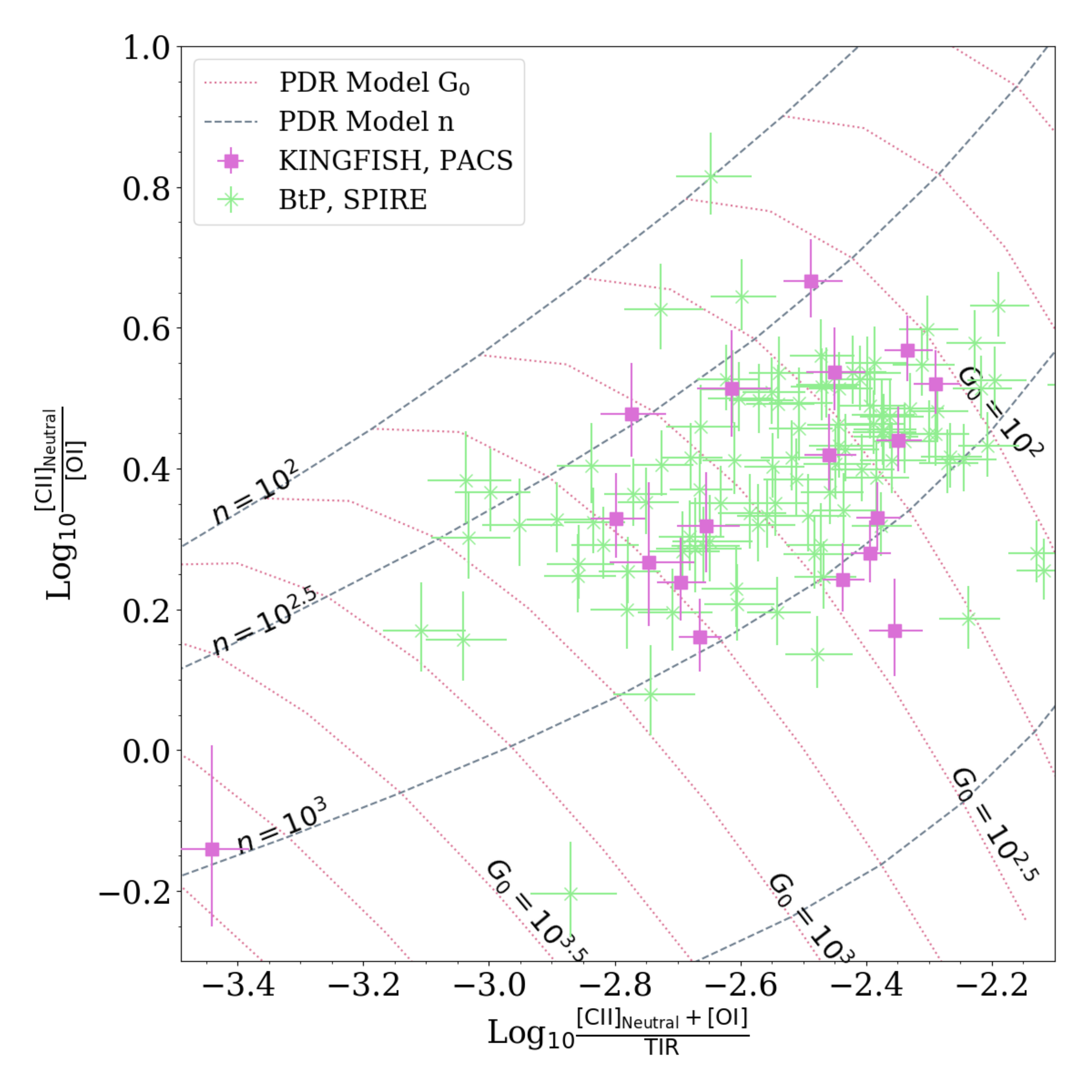}
\caption{This plot illustrates the PDR models used to determine $n_H$ for the neutral ISM.  The ratio of the \CII\ emission from the neutral ISM and the \OI~63~\micron\ line emission is plotted against the sum of the luminosity of these two lines divided by the TIR luminosity, a proxy for photoelectric heating efficiency.  Over-plotted are the $n_H$ and $G_0$ measurements determined by the PDR models of \citet{Kaufman2006, Pound2008}. }
\label{fig:pdrmodel}
\end{figure}

The density of the neutral phases were determined using PDR models.  We approximate \CII$_{\rm{Neutral}}$ emission as originating predominantly from PDRs, where we define PDRs as regions where FUV light dominates the heating and photochemistry within the ISM.  Working with this approximation, the PDR models of \cite{Kaufman2006, Pound2008} are ideal tools to estimate the properties of the neutral ISM in the regions included in this sample.  These models determine the density and FUV radiation intensity ($G_0$) by comparing the \CII~158~\micron\ and \OI~63~\micron\ line strengths, two of the predominant cooling channels in PDRs, as well as the sum of these two lines, to the total infrared emission, a measurement of the heating in PDRs.  The models of hydrogen nucleus density ($n_H$) and $G_0$ are over--plotted on measurements of the \CII~158~\micron~/~\OI~63~\micron\ vs the (\CII$_{\rm{Neutral}}$+\OI)/TIR$_{\rm{Neutral}}$ in Figure~\ref{fig:pdrmodel}. The \CII\ emission and the TIR luminosity from only the neutral phases were included for this analysis.  This method is limited by the effects of self--absorption and attenuation of the shorter wavelength \OI~63~\micron\ line, adding some inherit uncertainty to the measurements of $n_H$ obtained for the neutral ISM.

Using these PDR models, we find the regions in our sample cover a range of $G_0$ from 40--10$^4$, with a median value of $250$ and a range of $n_{\rm{H}}$ from 20--10$^{3.5}$ cm$^{-3}$ with a median value of $500$.  These results are what is expected for typical PDRs, and agree well with the work done in \citet{Malhotra2001, Croxall2012}.

\section{Thermalization and the \CII\ Deficit}
\label{sec:therm}

\subsection{Observations of the [CII] Deficit}
\begin{figure*}
\includegraphics[width=\linewidth]{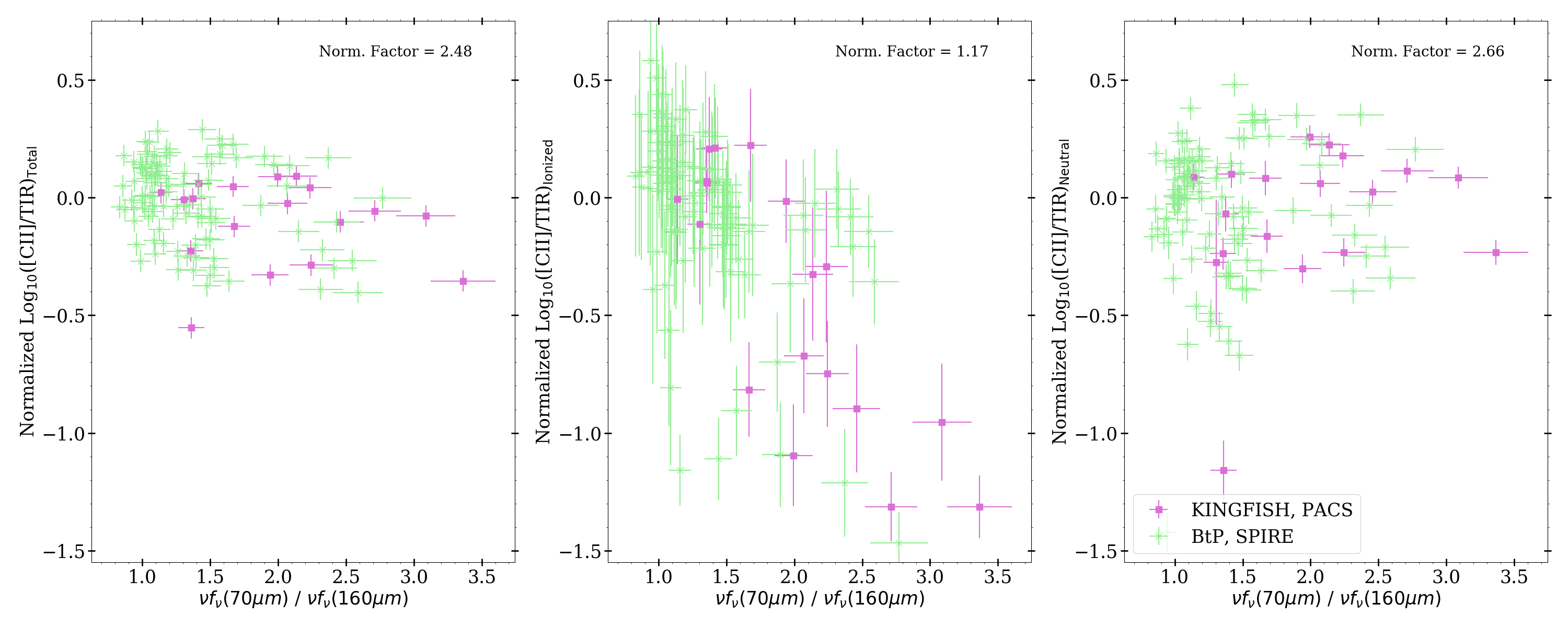}
\caption{\textit{Left}: The undivided \CII/TIR measurements are plotted against $\nu f_{\nu}(70 \mu \rm{m}) / \nu f_{\nu}(160 \mu \rm{m})$ flux ratios to show the moderate deficit behavior observed in the galaxies in this sample.  \textit{Middle}: The isolated ionized phase \CII/TIR measurements plotted against $\nu f_{\nu}(70 \mu \rm{m}) / \nu f_{\nu}(160 \mu \rm{m})$.  Notice the steeper negative slope indicating the strong deficit behavior in the ionized phase. \textit{Right}: The isolated neutral phase \CII/TIR measurements plotted against $\nu f_{\nu}(70 \mu \rm{m}) / \nu f_{\nu}(160 \mu \rm{m})$.  No deficit behavior is observed when only the neutral phase is considered.  This builds on the work performed in \citet{Sutter2019}.  All \CII/TIR ratios have been normalized by adding the absolute value of the average Log$_{10}$\CII/TIR ratio for the given phase to highlight the different trends in each phase.  For clarity the normalization factors are seen at the top of each panel.}
\label{fig:deftrad}
\end{figure*}

In order to characterize the \CII\ deficit behavior observed in this sample, the \CII/TIR measurements are plotted against the $\nu f_{\nu}(70 \mu \rm{m}) / \nu f_{\nu}(160 \mu \rm{m})$ values in Figure~\ref{fig:deftrad}.  The $\nu f_{\nu}(70 \mu \rm{m}) / \nu f_{\nu}(160 \mu \rm{m})$ flux ratio is an indicator of dust temperature, and therefore star formation activity, and is frequently used to show the effects of the \CII\ deficit \citep[e.g.][]{Croxall2012, DiazSantos2013, Malhotra1997}.  The left panel of this Figure shows the \CII/TIR ratio before any divisions by ISM phase are performed to either the \CII\ or TIR luminosities.  There is a slight decreasing trend with $\nu f_{\nu}(70 \mu \rm{m}) / \nu f_{\nu}(160 \mu \rm{m})$, indicating the regions in this sample experience moderate \CII\ deficit behavior.  The middle panel of Figure~\ref{fig:deftrad} displays the normalized isolated ionized phase \CII/TIR measurements and the right panel shows the  normalized isolated neutral phase \CII/TIR measurements.  All data has been normalized by dividing each point by the average value of \CII/TIR for that phase.  There is a clear decreasing trend between $\nu f_{\nu}(70 \mu \rm{m}) / \nu f_{\nu}(160 \mu \rm{m})$ and (\CII/TIR)$_{\rm{Ionized}}$.  This observed decreasing trend in the ionized phases of the ISM, along with the lack of a trend measured in the neutral phases (as shown in the right panel of Figure~\ref{fig:deftrad}) suggest that the cause of the slight decreasing trend for the star--forming regions included in this study is occurring in the ionized phases of the ISM.  For this reason, we focus our search for causes of the \CII\ deficit on trends we measure in the isolated ionized ISM phase \CII\ and TIR luminosities.

\subsection{Indications of Thermalization}
\label{sec:therm_obs}
\begin{figure*}
\includegraphics[width=\linewidth]{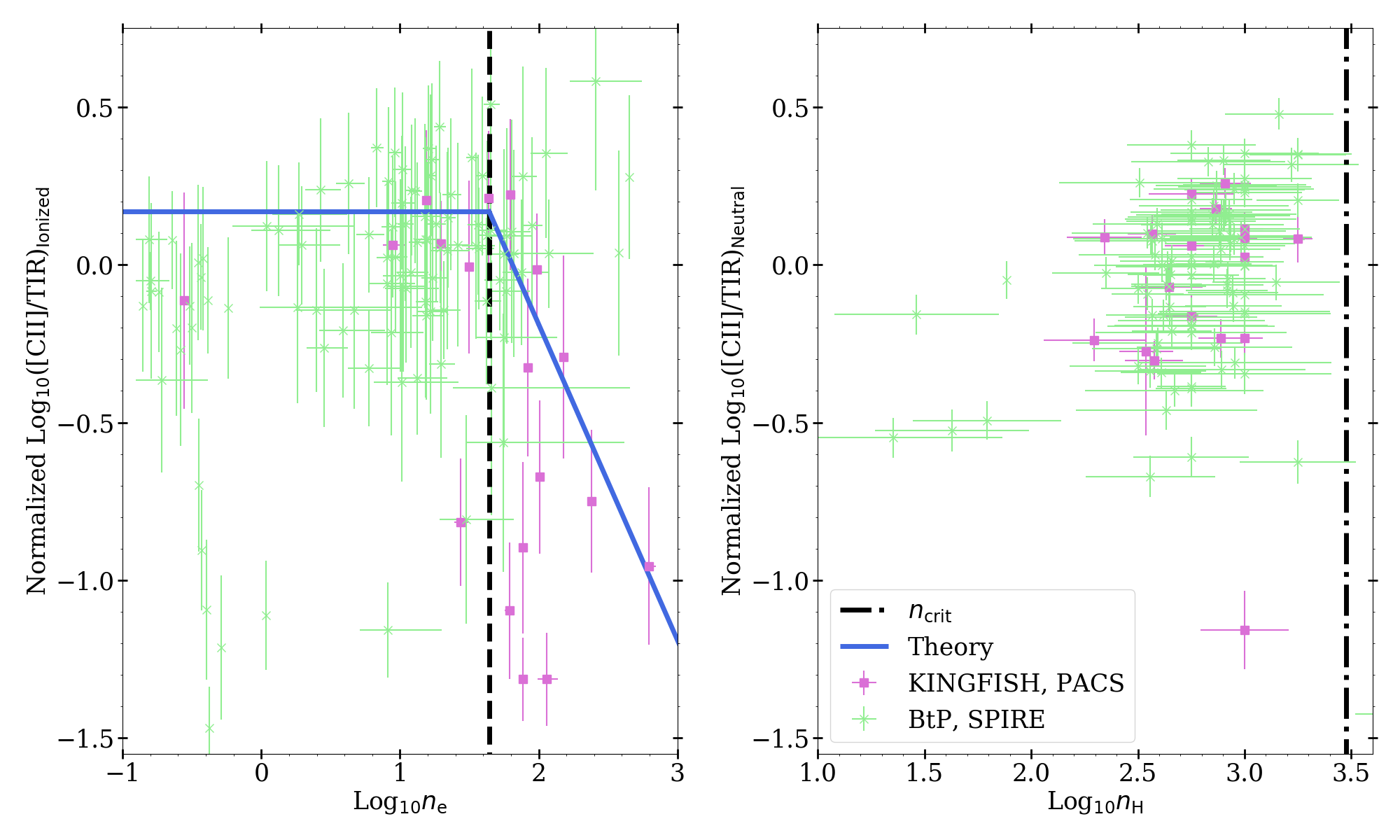}
\caption{The isolated--ISM phase ratios of \CII~/~TIR luminosity for the ionized (left) and neutral (right) phases of the ISM plotted against the electron number density in the ionized phase and the hydrogen atom number density for the neutral phase. The differing number densities were used as electrons are the primary source of collisional excitation for the \CII\ line in the ionized phases of the ISM, whereas hydrogen atoms are the primary source of collisional excitations of the \CII\ line in PDRs.  The black dotted and dashed lines represent the critical density of the \CII\ line for collisions with electrons (50 cm$^{-3}$) and hydrogen atoms (3$\times10^3$cm$^{-3}$) respectively \citep{Malhotra2001}.  The critical density for molecular hydrogen is a factor 1.4 higher.  The solid line in the left panel represents the median value of Log$_{10}$(\CII/TIR)$_{\rm{Ionized}}$ for regions with densities below the critical density.  The same normalization used in Figure~\ref{fig:deftrad} is applied here.}
\label{fig:moneymoney}
\end{figure*}

One finding from this novel approach of tracking the \CII\ deficit across the isolated ISM phases is the strong dependence of the observed deficit in the ionized phases of the ISM on the electron density relative to the critical density for the \CII\ line.  This is shown in Figure~\ref{fig:moneymoney}, where the \CII~/~TIR ratios for the isolated ionized and neutral ISM components are plotted against the electron number density for the ionized ISM, where free electrons are the primary source of collisional excitation of the \CII\ line, and the hydrogen nucleus number density for the neutral ISM. 
Hydrogen atoms are the primary source of collisional excitation of the \CII\
line at low column densities and molecular hydrogen dominates at the higher column
densities \citep[e.g.][]{HerreraCamus2015}.  The vertical dashed and dotted lines represent the critical electron and hydrogen atom number densities of \CII, respectively \citep{Goldsmith2012}.  These critical densities are determined assuming a temperature of 8000~K for the ionized phase and 100~K for the neutral phase.  The critical density for collisions with
molecular hydrogen is a factor ~1.4 higher than for collisions with hydrogen atoms \citep{Wiesenfeld2014}.  We see that in the ionized phases of the ISM, the \CII~/~TIR ratio plummets when the electron number density reaches the critical density.  This suggests that the thermalization of the \CII\ line is a factor in the observed \CII\ deficit in the ionized phases of the ISM.  In regions with densities above the critical density, increased heating of the dust and gas from young stars will increase the TIR luminosity, without increasing the \CII\ emission strength.
 
In the left panel of Figure~\ref{fig:moneymoney}, the sharp decrease along $n_{\rm crit}$ is primarily observed in the KINGFISH regions, represented by the pink squares.  This is likely a reflection of the selection of bright infrared regions for the KINGFISH observations.  Alternatively, the wider field of view maps obtained by the BtP survey cover the more quiescent areas within the ISM, where thermalization of the \CII\ is not occurring.  Although the KINGFISH regions cover areas of more intense star formation, none of the regions included in this sample probes the most extreme conditions where the largest deficits have been observed, as the BtP and KINGFISH sample cover a modest range of $L($TIR$)$ of $1\times10^8 - 7\times10^{11} L_{\odot}$. 

Despite an overall clear match between the theoretical values and observed data, there are a small subset of BtP regions below $n_{\rm{crit}}$ that have lower than predicted \CII/TIR (see green crosses in the lower right on the left panel of Figure~\ref{fig:moneymoney}) as well as two BtP regions above $n_{\rm{crit}}$ that are higher than expected (see green crosses in the upper right on the left panel of  Figure~\ref{fig:moneymoney}).  As all the BtP regions that fall far below the theoretical predictions have larger than average $f_{\rm{HII}}$ values, errors in how $f_{\rm{HII}}$ is determined are likely driving this discrepancy for some of these more quiescent BtP regions.  On the other hand, the points above the critical density and theoretical predictions have higher fractions of \CII\ emission from the ionized phases, which drives the (\CII/TIR)$_{\rm{Ionized}}$ values higher.  This could again be caused by a difference in the observed region and the assumptions made by modelling each region as an ionized \HII\ region surrounded by a neutral PDR.

It is important to note that for these samples, the declining \CII~/~TIR values as a function of density only occurs in the ionized ISM.  However, it is possible that the measurements of (\CII/TIR)$_{\rm{Neutral}}$ could also be effected by thermalization, but our simplistic models of the neutral medium having a temperature of 100 K and therefore a critical density of 3000 cm$^{-3}$ could be obscuring any effect thermalization has on the (\CII/TIR)$_{\rm{Neutral}}$ measurements. 

\subsection{Theoretical Predictions of Thermalization}
\label{sec:theory}
The observations discussed in Section~\ref{sec:therm_obs} suggest that thermalization of the \CII~158~\micron\ line is playing a role in the \CII\ deficit observed in the ionized ISM. This conclusion is supported by a theoretically derived model of the expected TIR and \CII\ luminosities, as elucidated in the following discussion.  Below the critical density, the \CII~158~\micron\ luminosity from \HII\ regions should be equal to the number of collisional excitations of C$^+$ multiplied by the energy produced by each transition:
\begin{equation}
    L( \mathrm{[CII]} , n < n_{\rm{\rm crit}}) = \frac{4}{3} \pi R^3 n_e n_{\rm{C^+}} \gamma_{\rm [CII]} E_{158}
\end{equation}
where $\gamma_{[CII]}$ is the rate coefficient for collisional excitations, $R$ is the radius of the \HII\ region, $E_{158}$ is the energy of a single transition, and $n_{\rm{e}}$ and $n_{\rm{C^+}}$ are the electron and singly--ionized carbon densities, respectively.  Above the critical density, the \CII\ luminosity will no longer depend on $n_e$, and can therefore be modelled as:
\begin{equation}
    L( \mathrm{[CII]} , n > n_{\rm{\rm crit}}) = \frac{4}{3} \pi R^3 n_{\rm{[CII]}} \gamma_{\rm [CII]} E_{158}.
\end{equation}
Regardless of density, the TIR luminosity from \HII\ regions can be modelled as:
\begin{equation}
    L(TIR) = N_{\rm{Ly}} E_{UV} f_{IR}
\end{equation}
where $N_{\rm{Ly}}$ is the number of Lyman continuum photons, $E_{UV}$ is the average energy of the Lyman continuum photons, and $f_{IR}$ is the fraction of energy from the Lyman continuum photons that is converted to dust heating.  Using Str\"{o}mgren conditions, $N_{\rm{Ly}}$ is:
\begin{equation}
    N_{\rm{Ly}} = \frac{4}{3} \pi R^3 n_e^2 \alpha
\end{equation}
where $\alpha$ is the recombination rate coefficient, and is approximately equal to $3\times10^{-13}$cm$^3$s$^{-1}$.  Using these predictions for the \CII\ and TIR luminosity for the ionized phase of the ISM, we can estimate the \CII/TIR ratio as a function of $n_e$. 
Below the critical density, \CII/TIR simplifies to:
\begin{equation}
\label{eq:def1}
    \frac{\mathrm{[CII]}}{\mathrm{TIR}} = \frac{ n_{\rm{[CII]}} \gamma_{\rm [CII]} E_{158}}{n_e \alpha E_{UV} f_{IR}}.
\end{equation}
If we assume $n_{\rm{[CII]}} \approx 1\times10^{-4} n_{\rm{e}}$ based on the Galactic carbon abundances \citep[1.6 $\times 10^{-4}$ carbon atoms per hydrogen atom, ][]{Sofia2004}, we can further simplify Equation~\ref{eq:def1} to:
\begin{equation}
    \frac{\mathrm{[CII]}}{\mathrm{TIR}} = \frac{ 1\times10^{-4} \gamma_{\rm [CII]} E_{158}}{\alpha E_{UV} f_{IR}}.
\end{equation}

Using $\alpha = 3\times10^{-13}$ cm$^3$s$^{-1}$, $\gamma_{[CII]} = 3.8\times10^{-7}$ \citep[][]{Tielens2005}, $E_{UV}$ of 15eV, and $f_{IR} = 0.5$ as suggested by \citet{Inoue2001}, we find:
\begin{equation}
   \frac{\mathrm{[CII]}}{\mathrm{TIR}} = 0.13 
\end{equation}
for the ionized phases of the ISM, when $n_{\rm{e}} \leq n_{\rm crit}$.  This value is displayed as a blue, horizontal line on the left panel of Figure~\ref{fig:moneymoney}, and is fairly consistent with the observed $\frac{\mathrm{[CII]}}{\mathrm{TIR}}$ values. As $f_{IR} = 0.5$ is a lower limit for the fraction of UV light reprocessed by dust, it is likely that the slight difference in the theory line and measured values is due to differences in this fraction across our sample \citep{Inoue2001}.

Above the critical density, we expect the \CII\ luminosity should no longer be proportional to $n_{\rm{e}}$.  The predicted $\frac{\mathrm{[CII]}}{\mathrm{TIR}}$ above $n_{\rm crit}$ is then:
\begin{equation}
    \frac{\mathrm{[CII]}}{\mathrm{TIR}} = \frac{0.13 n_{\rm crit}}{n_e}.
\end{equation}
The theoretical predictions described in this section are displayed as a solid blue line on the left-hand panel of Figure~\ref{fig:moneymoney}.  It is likely that the theoretical predictions lie below the observed values due to the possibility that some of the \CII\ emission from ionized phases originates in the diffuse ionized ISM, while these predictions only account for the \CII\ emission from \HII\ regions.  Emission from the diffuse ionized ISM is also expected to play a greater role in the BtP data points, which cover the areas surrounding the star--forming regions included in the KINGFISH study.  The discrepancy between the theoretical curve and the observed measurements could also be driven by underestimates in our prediction for $f_{IR}$.  As \citet{Inoue2001} focuses on star--forming regions in local galaxies and states $f_{IR} = 0.5$ as an upper limit, it is likely that differences in $A_V$ or other conditions in the star--forming regions targeted in this study could lead to variations in the amount of UV light absorbed by dust, increasing the value of $f_{IR}$.

\section{Discussion and Conclusion}
\label{sec:con}

\begin{figure*}
\includegraphics[width=\linewidth]{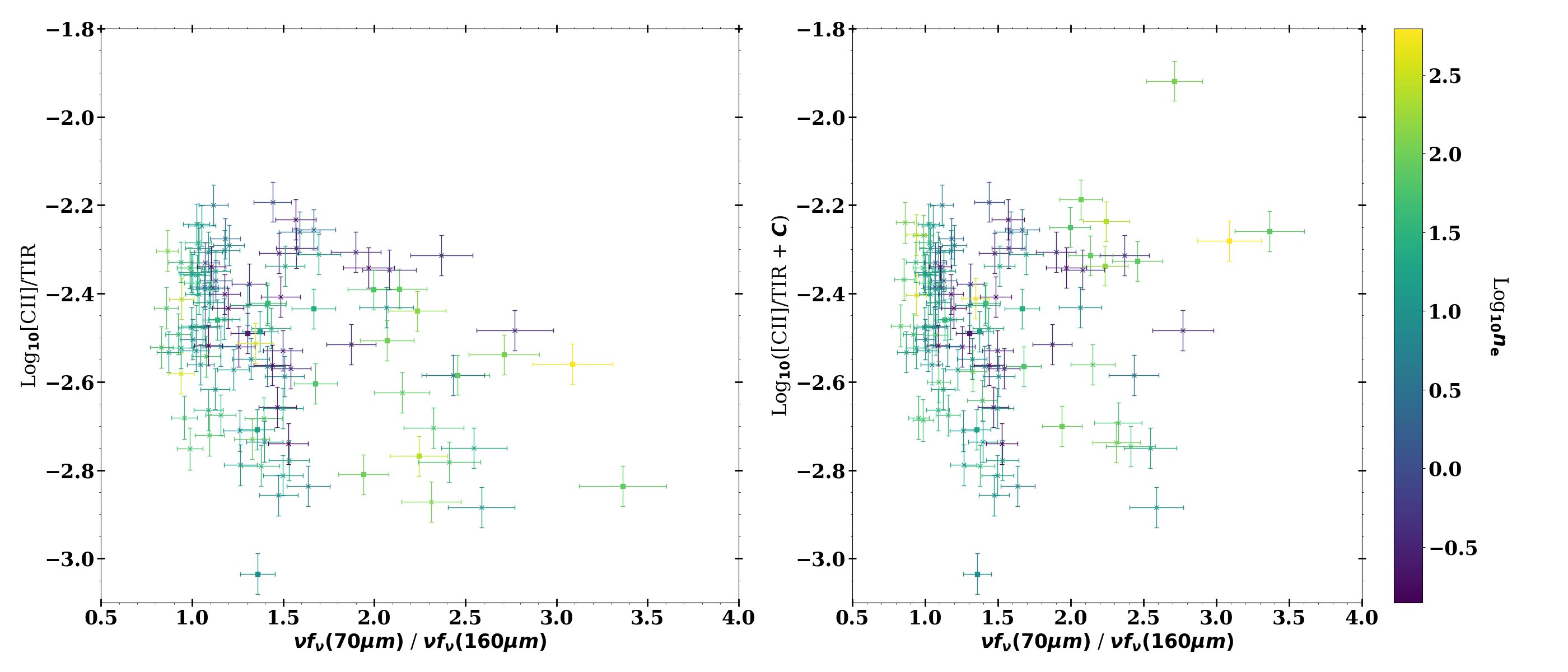}

\caption{Left Panel: The total \CII~/~TIR measurements for the regions in this sample color-coded by  Log$_{10} n_{\rm{e}}$ against $\nu f_{\nu}(70\mu \rm{m}) / \nu f_{\nu}(160\mu \rm{m})$, an indicator of dust temperature.  Filled square points represent KINGFISH regions while crosses represent the BtP regions.  Right Panel: Same as the left panel, with the \CII/TIR ratios corrected for the effects of thermalization through the addition of the correction factor $C$ described in Equation~\ref{eq:corr}, removing the observed deficit.}
\label{fig:ciidef}
\end{figure*}

As previous studies have found that typically $\approx 75\%$ of \CII\ emission originates in PDRs and other neutral phases of the ISM \citep{Croxall2017, Rigopoulou2013}, thermalization of \CII\ in the ionized ISM can only account for moderate overall deficit behavior.  For the normal star--forming galaxies in this work, the undivided \CII/TIR measurements showed only moderate deficit behavior.  Therefore, the \CII\ deficit observed in this sample can be explained by thermalization of the \CII\ line within the ionized phases of the ISM, despite the majority of \CII\ emission originating in the neutral ISM \citep[See Figure~\ref{fig:deftrad} or ][for further discussion]{Sutter2019}. Furthermore, by correcting for the effects of thermalization in the ionized phases of the ISM, the deficit effect can be greatly reduced.  This is shown by the differences in the left and right panels of Figure~\ref{fig:ciidef}.  The left panel shows the unmodified \CII~/~TIR measurements for the regions in this sample color coded by Log$_{10} n_{\rm{e}}$.  The total \CII~/~TIR measurements show a slight decline as a function of $\nu f_{\nu}(70\mu \rm{m}) / \nu f_{\nu}(160\mu \rm{m})$, a proxy for dust temperature.  This decrease is removed in the right panel of Figure~\ref{fig:ciidef}, where the \CII/TIR measurements have been corrected for the effects of thermalization in the ionized ISM.  This correction is made using the theoretical predictions described in Section~\ref{sec:theory}.  For regions above the critical density, the \CII/TIR values are corrected by adding the correction factor:
\begin{equation}
    \label{eq:corr}
    C= 0.13 (1.0 - \frac{ n_{\rm{\rm crit}}}{n_{\rm{e}}}) \times f_{\rm HII}
\end{equation}

Here, $C$ represents the difference between the theoretical expectation of non--thermalized \CII/TIR (0.13) and thermalized \CII/TIR ($\frac{0.13n_{\rm{\rm crit}}}{n_{\rm{e}}}$) from the ionized phases of the ISM.
The left-hand panel of Figure~\ref{fig:ciidef} plots Log$_{10}$(\CII/TIR + C) against the $\nu f_{\nu}(70\mu \rm{m}) / \nu f_{\nu}(160\mu \rm{m})$ measurements.  This correction for the effects of thermalization removes the observed decrease seen in the undivided \CII/TIR data, but still shows a large scatter, indicating that there are likely other physical processes effecting the \CII/TIR ratio in these galaxies.

Thermalization may also be the cause for deficits observed in other ionized gas tracers, like \NII\ and \OIII\ lines \citep{HerreraCamus2016, GraciaCarpio2011}, and therefore should be accounted for in studies of ionized gas in galaxies.  As the deficit is most prominent in the ionized gas, it is also possible that some of this behavior is due to increasing levels of doubly--ionized carbon.  With our current data set, it is difficult to disentangle how thermalization of \CII\ and a larger fraction of C$^{++}$ both contribute to the observed deficit in the ionized ISM.

It should be noted that despite the evidence presented here, thermalization in the ionized phases of the ISM cannot be the only contributor to the \CII\ deficit.  This is especially true when considering the high--z galaxies with especially pronounced deficit behavior.  For example, the z~$\sim$~2--6 galaxies observed in \citet{Rawle2014, Capak2015, DeBreuck2014, Gallerani2012, Malhotra2017} show \CII/TIR ratios 10 to 100 times lower than those observed in the local star--forming galaxies included in this work.  As the majority of \CII\ emission has been found to arise from the neutral phases of the ISM \citep{Croxall2017, Pineda2013, Abdullah2017, Sutter2019, Cormier2019, Bigiel2020}, the effects of thermalization solely within the ionized phases of the ISM can only contribute to moderate deficit behavior.  This suggests that although thermalization of the \CII\ line within the ionized phases of the ISM may play an important role in the \CII\ deficit, especially over the kpc--scale star--forming regions included in this work, it cannot be the only factor consider to drive this behavior across a wide spectrum of spatial resolutions and galaxy conditions.

Although thermalization of \CII\ in \HII\ regions cannot completely explain the large deficits observed in some starbursting galaxies \citep{Smith2017} or the \CII~158~\micron\ emission behavior in low metallicity dwarfs where little \CII\ emission originates in ionized phases \citep{Cormier2015}, the sharp decrease in (\CII~/~TIR)$_{\rm{Ionized}}$ along $n_{\rm{\rm crit}}$ is an intriguing result that will help us decode the behavior of the \CII\ deficit across the range of galaxies in which it has been observed.  Further studies of these isolated ISM phases will help define what other process could affect the strength of the \CII\ line, and will have important implications for the use of this line as a tracer of conditions in the high--redshift universe.  

\section*{Acknowledgements}
This work was supported by NASA Headquarters under the
NASA Earth and Space Science Fellowship Program,
Grant \#80NSSC18K1107, as the Wyoming NASA Space
Grant Consortium, NASA Grant \#NNX15AI08H. Herschel is an ESA space observatory with science instruments provided by European-led Principal Investigator
consortia and with important participation from NASA.
IRAF, the Image Reduction and Analysis Facility, has
been developed by the National Optical Astronomy Observatories and the Space Telescope Science Institute.
\section*{Data Availability}

All data are available through IRSA with the exception of the SPIRE spectroscopy which is available upon request.



\bibliographystyle{mnras}
\bibliography{main} 




\bsp	
\label{lastpage}
\end{document}